\documentstyle[amssymb,aps,preprint,12pt]{revtex}

\begin{document}
\author{J. C. Stenerson}
\title{A Simple RevTeX Article}
\date{}

\begin{center}
{\Large Plasma dispersion of multisubband electron systems over liquid helium%
}

\medskip 

Sviatoslav S. Sokolov

{\small Departamento de F\'{i}sica, Universidade Federal de S\~{a}o Carlos,
13565-905 S\~{a}o Carlos, S\~{a}o Paulo, Brazil and B. I. Verkin Institute
for Low Temperature Physics and Engineering, National Academy of Sciences of
Ukraine, 61164 Kharkov, Ukraine}

\smallskip 

Nelson Studart

{\small Departamento de F\'{i}sica, Universidade Federal de S\~{a}o Carlos,
13565-905 S\~{a}o Carlos, S\~{a}o Paulo, Brazil}

\bigskip
\end{center}

\begin{quote}
Density-density response functions are evaluated for nondegenerate
multisubband electron systems in the random-phase approximation for
arbitrary wave number and subband index. We consider both
quasi-two-dimensional and quasi-one-dimensional systems for electrons
confined to the surface of liquid helium. The dispersion relations of
longitudinal intrasubband and transverse intersubband modes are calculated
at low temperatures and for long wavelengths. We discuss the effects of
screening and two-subband occupancy on the plasmon spectrum. The
characteristic absorption edge of the intersubband modes is shifted
relatively to the single-particle intersubband separation and the
depolarization shift correction can be significant at high electron
densities.

\smallskip PACS number(s): 73.20.Dx; 73.20.Mf; 73.90.+f
\end{quote}

\smallskip

\section{Introduction}

Two-dimensional electron systems (2DES) over the liquid helium surface have
been intensively studied for a long time.\cite{andrei} More recently, it was
possible to confine these surface electrons (SE) in reduced geometries
creating also one-dimensional (1D) systems.\cite{q1d-exp}. Both systems
provide a nearly ideal laboratory for studying collective phenomena in the
electron plasma in lower dimensions because the cleanness of the helium
surface restricts SE scattering mechanisms to those with helium atoms in the
vapor phase, which predominates at $T>1$ K, and with surface oscillations
(ripplons) at lower temperatures. Furthermore both scattering mechanisms
become ineffective with lowering temperature and can be discarded at $%
T\lesssim 0.1$ K. In such a regime, collective effects in low-dimensional
electron systems due to the Coulomb interaction can be investigated ignoring
the interaction with scatterers. Another important feature of these systems
is the accessible range of SE densities which is limited to $n_{s}\lesssim
10^{9}$ cm$^{-2}$ (for bulk helium). As a consequence, the 2D Fermi energy $%
\varepsilon _{F}\lesssim 10^{-2}$ K and SE behave like nondegenerate
low-dimensional systems differing in many aspects from its quantum
counterpart realized in semiconductor structures.\cite{an-fowl-stern.82}

As it is well known, the collective excitation spectrum depends crucially on
the way the particles are confined. For instance, for longitudinal plasma
oscillations of the 2DES, the spectrum $\omega _{2D}(q)\sim q^{1/2}$ is in
contrast to the 3D situation in which one has a optical mode starting from
the plasma frequency. This is a consequence of the fact that the screening
is incomplete in 2D because there are electromagnetic fields in the vacuum
surrounding the plane and many-body effects play important role in
describing the properties of the 2DES. On the other hand, the longitudinal
plasmon mode in the 1DES case is $\omega _{1D}(q_{x})\sim q_{x}\ln
(q_{x}\ell $), where $\ell $ is some characteristic length of the system. In
these cases, we have assumed that only the lowest subband, for electron
motion along the direction perpendicular to the electron sheet, is occupied.
This limiting case is achieved when the Bolztmann factor $\exp (-\Delta
_{21}/T)\ll 1$, where $\Delta _{21}=\Delta _{2}-\Delta _{1}$ is the energy
gap between the lowest (1) and the first-occupied (2) subband, and the
occupation of higher subbands is negligible. Otherwise, the multisubband
nature of low-dimensional electron systems $-$ hereafter referred as
quasi-2D(1D)ES $-$ cannot be discarded when the temperature is comparable
with $\Delta _{21}$ and population effects of higher subbands cannot be
ignored.

In this paper, we address the problem of plasmon spectrum in Q2DES and Q1DES
over the surface of liquid helium. We use the many-body dielectric formalism
within the random-phase approximation. In this approach, the mode spectrum
is obtained from the roots of a determinantal equation for the dielectric
function. At first glance, we note that the multisubband character of these
systems allows the existence of transverse modes of plasma oscillations in
the direction normal to that of unconfined electron motion.

We adopt a two-subband model where the bare electron-electron interaction is
evaluated using subband wave functions found by the variational method for
the Q2DES and taken as the harmonic-oscillator functions for the parabolic
confinement in the Q1DES. We limit ourselves to the case of low enough
temperatures which allows us to disregard the coupling of plasma
oscillations with ripplon modes. We do not also consider the possible
transition of the electron system to the ordered state where the
electron-ripplon interaction can strongly modify the mode spectrum. \cite
{fish-halp-platzm,mon-shik.82,sok-stud.99}

\section{Theoretical approach}

The main theoretical approach to the study of plasma oscillations in
multisubband low-dimensional charge system is based on many-body dielectric
formalism using the generalized dielectric function

\begin{equation}
\epsilon _{nn^{\prime },mm^{\prime }}(\omega ,q)=\delta _{nn^{\prime
}}\delta _{mm^{\prime }}-V_{nn^{\prime },mm^{\prime }}(q)\Pi _{mm^{\prime
}}(\omega ,q)  \label{1}
\end{equation}
where $\Pi _{mm^{\prime }}(\omega ,q)$ is the density-density response
function, $\delta _{nn^{\prime }}$ is the Kronecker symbol and $%
V_{nn^{\prime },mm^{\prime }}(q)$ is the matrix element of
Fourier-transformed Coulomb interaction averaged over wave functions of
subbands with indices $n$, $n^{\prime }$, $m$, and $m^{\prime }$ equal to $%
1, $ $2,$ $3...$. The dielectric function $\epsilon _{nn^{\prime
},mm^{\prime }}(\omega ,q)$ depends both on the frequency and the wave
numbers $q$ for the Q2DES and $q_{x}$ for the Q1DES.

In the random-phase approximation (RPA), we assume that the electron system
responds to external perturbations as a noninteracting system and we take $%
\Pi _{mm^{\prime }}(\omega ,q)=\Pi _{mm^{\prime }}^{0}(\omega ,q)$, where
the free polarizability function is written as 
\begin{equation}
\Pi _{mm^{\prime }}^{0}(\omega ,q)=\sum_{{\bf k},\sigma }\frac{f_{0}(E_{{\bf %
k}}+\Delta _{m})-f_{0}(E_{{\bf k+q}}+\Delta _{m^{\prime }})}{\hbar \omega
+E_{{\bf k}}+\Delta _{m}-E_{{\bf k+q}}-\Delta _{m^{\prime }}+i\delta }.
\label{2.1}
\end{equation}
Here $E_{{\bf k}}=\hbar ^{2}k^{2}/2m,$ where $m$ is the electron mass, $%
\delta $ is infinitesimal positive, and $\sigma $ is spin index. For
classical systems, the distribution function $f_{0}(E_{{\bf k}}+\Delta
_{n})=\exp \left[ -\left( E_{{\bf k}}+\Delta _{n}\right) /T\right] $ and is
normalized by the condition $\sum_{n,_{{\bf k}},\sigma }f_{0}(E_{{\bf k}%
}+\Delta _{n})=N$ where $N$ is the number of particles.

Using the dielectric function given by Eq. (\ref{1}), Vinter \cite{vinter.77}
and Das Sarma \cite{sarma.84} have studied many-body effects in the
degenerate Q2DES. Das Sarma and co-workers \cite
{sarma-lai.85,li-sarma.90,li-sarma.91}, Hu and O$^{\prime }$Connell \cite
{hu-oconnell.90} and Hai {\it et al}. \cite{hai97} extended these studies to
plasma oscillations in degenerate Q1D multisubband system whereas Sokolov
and Studart \cite{sok-stud.98} approach the problem in the classical regime.

The well-known bare electron-electron potential is given by 
\begin{equation}
V_{nn^{\prime },mm^{\prime }}^{2D}(q)=\int_{0}^{\infty }\int_{0}^{\infty
}dzdz_{^{\prime }}\psi _{n}(z)\psi _{n^{\prime }}(z)v^{2D}(q)\psi
_{m}(z^{\prime })\psi _{m^{\prime }}(z^{\prime }),  \label{1.2}
\end{equation}
and 
\begin{equation}
V_{nn^{\prime },mm^{\prime }}^{1D}(q_{x})=\int_{-\infty }^{\infty
}\int_{-\infty }^{\infty }dydy^{\prime }\varphi _{n}(y)\varphi _{n^{\prime
}}(y)v^{1D}(q_{x})\varphi _{m}(y^{\prime })\varphi _{m^{\prime }}(y^{\prime
}),  \label{1.3}
\end{equation}
where $v^{2D}(q)=2\pi \widetilde{e}^{2}/Sq$, [$v^{1D}(q_{x})=2(\widetilde{e}%
^{2}/L_{x})K_{0}(|q_{x}||y-y^{\prime }|)$ is the Coulomb potential, $S$ [$%
L_{x}$] the area [length] of the system, and $\psi _{n}(z)$ [$\varphi _{n}(y)
$] denotes the $n$-th subband wave functions for the Q2DES [Q1DES]. Here $%
\widetilde{e}=\left[ 2e^{2}/(1+\varepsilon )\right] $, with $\varepsilon $
the helium dielectric constant, is the effective charge taking substrate
effects into account.

\section{Plasmon spectrum}

The dispersion relations for collective modes for a multisubband system are
found from the roots of the determinantal equation

\begin{equation}
\det |\epsilon _{nn^{\prime },mm^{\prime }}(q,\omega )|=0.  \label{2}
\end{equation}
In principle, all the subbands should be considered in the above equation.
However an useful analytical solution is possible in a two-subband model. In
this case, Eq. (\ref{2}) splits into two independent equations 
\begin{mathletters}
\begin{equation}
1-V_{11,11}\Pi _{11,11}^{0}(\omega ,q)=0  \label{3a}
\end{equation}

\begin{equation}
1-V_{12,12}\left[ \Pi _{12}^{0}(\omega ,q)+\Pi _{21}^{0}(\omega ,q)\right]
=0.  \label{3b}
\end{equation}
Mode coupling appears only if one take into account higher subbands.
Equation (\ref{3a}) describes the longitudinal {\it intrasubband }plasma
oscillations whose dispersion law must coincide with that of 2DES or 1DES
with one-subband occupancy system whereas Eq. (\ref{3b}) gives the
dispersion law for transverse {\it intersubband }oscillations involving
transitions from the lowest to the second subband.

\subsection{Q2DES}

As it is well-known, SE on helium are trapped in the direction perpendicular
to the surface ($z$ direction) by a potential well due to image forces and a
holding electric field $E_{\bot }$. For $E_{\bot }=0,$ and the image
potential $V(z)=-\Lambda _{0}/z$, where $\Lambda _{0}=(e^{2}/4)(\varepsilon
-1)/(\varepsilon +1)$, and infinite potential barrier at the interface, the
solution of the Schr\"{o}dinger equation is given by \cite
{cole69,shikin70,nieto.99}

\end{mathletters}
\begin{equation}
\psi _{n}(z)=\frac{2\kappa _{0}^{3/2}z}{n^{5/2}}\exp \left( -\frac{\kappa
_{0}z}{n}\right) L_{n-1}^{(1)}\left( \frac{2\kappa _{0}z}{n}\right)
\label{5}
\end{equation}
where $\kappa _{0}=m\Lambda _{0}/\hbar ^{2}$ ($=3/(2\left\langle
z\right\rangle _{0})$, where $\left\langle z\right\rangle _{0}$ is the mean
electron distance from the plane) and $L_{n}^{(\alpha )}(x)$ are the
associated Laguerre polynomials. The energy subband is given by the
hydrogen-like spectrum $\Delta _{n}=\Delta _{0}/n^{2}$ where $\Delta
_{0}=\hbar ^{2}\kappa _{0}^{2}/2m$. If the pressing electric field $E_{\perp
}$ is turned on, there is no general analytical solution and we assume trial
wave functions corresponding to two lowest subbands ($n=1$ and $2$) of Eq. (%
\ref{5}) with variational parameters $\kappa _{1}$ and $\kappa _{2}$:\cite
{mon-shik-sok.81,sok.85} 
\begin{mathletters}
\begin{equation}
\psi _{1}(z)=2\kappa _{1}^{3/2}z\exp \left( -\kappa _{1}z\right) ,
\label{7a}
\end{equation}

\begin{equation}
\psi _{2}(z)=\frac{2\sqrt{3}\kappa _{2}^{5/2}}{\kappa _{12}}\left[ 1-\left( 
\frac{\kappa _{1}+\kappa _{2}}{3}\right) z\right] z\exp \left( -\kappa
_{2}z\right) ,  \label{7b}
\end{equation}
and subband energies:

\begin{equation}
\Delta _{1}=\frac{\hbar ^{2}\kappa _{1}^{2}}{2m}-\Lambda _{0}\kappa _{1}+%
\frac{3eE_{\perp }}{2\kappa _{1}},  \label{7c}
\end{equation}

\begin{equation}
\Delta _{2}=\frac{\hbar ^{2}\kappa _{2}^{2}}{6m}\left[ 1+\frac{6\kappa
_{2}^{2}}{\kappa _{12}^{2}}\right] -\frac{\Lambda _{0}\kappa _{2}}{2}\left[
1+\frac{2\kappa _{2}^{2}-\kappa _{1}\kappa _{2}}{\kappa _{12}^{2}}\right] +%
\frac{eE_{\perp }}{2\kappa _{2}}\left[ 1+\frac{4\kappa _{1}^{2}-\kappa
_{1}\kappa _{2}+\kappa _{2}^{2}}{\kappa _{12}^{2}}\right] .  \label{7d}
\end{equation}
Here $\kappa _{12}^{2}=\kappa _{1}^{2}-\kappa _{1}\kappa _{2}+\kappa
_{2}^{2}.$ If we define $\kappa _{1}(E_{\perp })=\eta _{1}\kappa _{0}$ and $%
\kappa _{2}(E_{\perp })=\eta _{2}\kappa _{0},$ one can find $\eta _{1}$ and $%
\eta _{2}$ as the roots of the system of equations given by 
\end{mathletters}
\begin{mathletters}
\begin{equation}
\eta _{1}^{3}-\eta _{1}^{2}-\left( \frac{\kappa _{\perp }}{\kappa _{0}}%
\right) ^{3}=0,  \label{8a}
\end{equation}

\[
\eta _{2}^{3}(\eta _{1}^{4}-2\eta _{1}^{3}\eta _{2}+15\eta _{1}^{2}\eta
_{2}^{2}-11\eta _{1}\eta _{2}^{3}+7\eta _{2}^{4})-\frac{3}{2}\eta
_{2}^{2}(\eta _{1}^{4}-4\eta _{1}^{3}\eta _{2}+10\eta _{1}^{2}\eta
_{2}^{2}-6\eta _{1}\eta _{2}^{3}+3\eta _{2}^{4}) 
\]

\begin{equation}
-\left( \frac{\kappa _{\perp }}{\kappa _{0}}\right) ^{3}(5\eta
_{1}^{4}-10\eta _{1}^{3}\eta _{2}+15\eta _{1}^{2}\eta _{2}^{2}-4\eta
_{1}\eta _{2}^{3}+2\eta _{2}^{4})=0  \label{8b}
\end{equation}
where $\kappa _{\perp }=(3meE_{\perp }/2\hbar ^{2})^{1/3}.$ For $E_{\perp }=0
$, Eqs. (\ref{8a}) and (\ref{8b}) reproduce the results, given by Eq. (\ref
{5}) and respective eigenenergies with $\eta _{1}=1$ and $\eta _{2}=0.5$.
The numerical values of the variational parameters as a function of the
pressing electric field are plotted in Fig. 1. We observe a rapid increase
at low field and an asymptotic linear behavior at larger fields. With these
values for $\eta _{1}$ and $\eta _{2}$, we depicted in Fig. 2, the field
dependence of the energy of the lowest-subband and the energy gap $\Delta
_{21}$.

Using Eqs. (\ref{1.2}), (\ref{7a}) and (\ref{7b}) one can calculate the
values of $V_{11,11}$ and $V_{12,12}$ up to second-order in the parameters $%
q/\kappa _{1}\ll 1$ and $q/(\kappa _{1}+\kappa _{2})\ll 1$ as 
\end{mathletters}
\begin{mathletters}
\begin{equation}
V_{11,11}=v^{2D}(q)\left[ 1-\frac{3q}{4\kappa _{1}}+\frac{3q^{2}}{4\kappa
_{1}^{2}}\right] ;  \label{10a}
\end{equation}

\begin{equation}
V_{12,12}=v^{2D}(q)\alpha (E_{\perp })\frac{q}{\kappa _{0}}\left[ 1-\frac{16q%
}{5(\kappa _{1}+\kappa _{2})}+\frac{7q^{2}}{(\kappa _{1}+\kappa _{2})^{2}}%
\right]  \label{10b}
\end{equation}
\smallskip with $\alpha (E_{\perp })=60\eta _{1}^{3}\eta _{2}^{5}/[(\eta
_{1}+\eta _{2})^{7}(\eta _{1}^{2}-\eta _{1}\eta _{2}+\eta _{2}^{2})].$ The
well-behaved form of $\alpha (E_{\perp })$ does not influence strongly $%
V_{12,12}$ because $\alpha (E_{\perp }=0)=0.146$ and $\alpha (E_{\perp })$
increases by increasing $E_{\perp }$ until reaches a maximum $\alpha _{\max
}=0.281$ near $E_{\perp }=0.3$ kV/cm and slightly decreases to $0.227$ at $%
E_{\perp }=3$ kV/cm.

For the Q2DES, the noninteracting density-density response function, Eq.(\ref
{2.1}), can be calculated in a straightforward way. The result is

\end{mathletters}
\begin{equation}
\Pi _{nn^{\prime }}^{0}(\omega ,q)=-\frac{N}{\hbar qu_{T}Z_{n}}\left[ \exp
\left( -\Delta _{n}/T\right) U\left( \zeta _{nn^{\prime }}^{(-)}\right)
-\exp \left( -\Delta _{n^{\prime }}/T\right) U\left( \zeta _{nn^{\prime
}}^{(+)}\right) \right]  \label{12}
\end{equation}
\smallskip where $\zeta _{nn^{\prime }}^{(\pm )}=\left[ \omega +\left(
\Delta _{n}-\Delta _{n^{\prime }}\right) /\hbar \right] /qu_{T}\pm \hbar
q/2mu_{T}$, with $u_{T}=\sqrt{2T/m}$ being the thermal velocity and $%
Z_{n}=\sum_{n}\exp \left( -\Delta _{n}/T\right) $. Similar general structure
of $\Pi _{nn^{\prime }}^{0}(\omega ,q)$ is found in the classical regime of
the electron gas in 3D case. \cite{fett-walecka.71} The function $U(\zeta )$
is given by the integral

\begin{equation}
U(\zeta )=\frac{1}{\sqrt{\pi }}\int_{-\infty }^{\infty }\frac{\exp (-y^{2})}{%
y-\zeta -i\delta }=-2\exp (-\zeta ^{2})\int_{0}^{\zeta }\exp (t^{2})dt+i%
\sqrt{\pi }\exp (-\zeta ^{2}).  \label{13}
\end{equation}
For $n=n^{\prime }$ and for small $q\ll (2m\omega /\hbar )^{1/2}$ Eq. (\ref
{12}) can be approximately expressed through $W(\zeta ),$ the well-known
function in the plasma theory, as \cite
{platz-tzoar.76,totsuji,mon.77,stud-hip.79}

\begin{equation}
\Pi _{nn^{\prime }}^{(0)}(\omega ,q)\simeq -\frac{N}{TZ_{n}}W\left( \frac{%
\omega }{qu_{T}}\right) \exp \left( -\frac{\Delta _{n}}{T}\right) .
\label{14}
\end{equation}
The function $W(\zeta )$ is connected with $U(\zeta )$ by the relation

\[
W(\zeta )=-(\partial U/\partial \zeta )/2=1+\zeta U(\zeta ). 
\]

Putting $\omega =\omega (q)-i\gamma _{q},$ and assuming $\omega
(q)/qu_{T}\gg 1$ and $\left| \omega (q)-\Delta _{21}/\hbar \right|
/qu_{T}\gg 1,$ we obtain, in the two-subband model, the dispersion relation
for longitudinal intrasubband mode frequencies ($\omega _{l}$) and damping ($%
\gamma _{l}$): 
\begin{mathletters}
\begin{equation}
\omega _{l}^{2}(q)=\omega _{2D}^{2}(q)\left[ 1+\frac{3q}{k_{D}}-\frac{3q}{%
4\kappa _{1}}\right] ,  \label{15a}
\end{equation}

\begin{equation}
\gamma _{l}(q)=\sqrt{\pi }\frac{\omega _{l}^{4}(q)}{(qu_{T})^{3}}\exp \left[
-\frac{\omega _{l}^{2}(q)}{(qu_{T})^{2}}\right] ,  \label{15b}
\end{equation}
where $\omega _{2D}^{2}(q)=(2\widetilde{e}^{2}/ma^{2})q$ and $k_{D}=2%
\widetilde{e}^{2}/Ta^{2}$ is the 2D Debye wave number and $a=(\pi
n_{s})^{-1/2}$ is the mean interelectron spacing. One can see, from Eq. (\ref
{15a}), that the first two-terms of the real part of the longitudinal branch
are the same as in the classical 2DES\cite{platz-tzoar.76,totsuji,mon.77}.
The extra term ($3q/4\kappa _{1}$) comes from the effect of the layer
thickness, because $\kappa _{1}$ is related to the distance of the electron
in the lowest subband from the plane, and lowers the dispersion relation at
small wavelengths. From Eq. (\ref{15a}), we also observe the competition
between screening and layer thickness effects on the Q2DES. The Landau
damping $\gamma _{l}(q)$ of the plasma oscillation is practically the same
as in the 2DES.

For transverse intersubband modes, we obtain the frequencies ($\omega _{t}$)
and damping ($\gamma _{t}$): 
\end{mathletters}
\begin{mathletters}
\begin{equation}
\omega _{t}^{2}(q)\simeq \omega _{21}^{2}+2\omega _{21}\omega _{2D}^{(sh)} 
\left[ 1+\frac{\left( 1+3\omega _{2D}^{(sh)}/2\omega _{21}\right) Tq^{2}}{%
m\left( \omega _{2D}^{(sh)}\right) ^{2}}-\frac{16q}{\kappa _{1}+\kappa _{2}}%
\right] ,  \label{16a}
\end{equation}

\begin{equation}
\gamma _{t}(q)\simeq \sqrt{\pi }\frac{\left[ \omega _{t}(q)-\omega _{21}%
\right] ^{2}}{qu_{T}}\exp \left[ -\frac{\left[ \omega _{t}(q)-\omega _{21}%
\right] ^{2}}{(qu_{T})^{2}}\right] .  \label{16b}
\end{equation}
where $\omega _{21}=\Delta _{21}/\hbar $ and $\omega _{2D}^{(sh)}=2\alpha
(E_{\perp })\widetilde{e}^{2}/\hbar \kappa _{0}a^{2}$. As one can see the
transverse plasmon mode spectrum, given by Eq. (\ref{16a}), is quite
different of the longitudinal branch and has a gap at $q=0.$ The frequency
of the characteristic absorption edge is shifted, relatively to the
frequency $\omega _{12}$ of the intersubband transition, shown in Fig. 2, by 
$\Delta \omega =\sqrt{\omega _{21}^{2}+2\omega _{21}\omega _{2D}^{(sh)}}%
-\omega _{21}$, which is the manifestation of the depolarization shift
effect in transverse oscillations of the many-body system. \cite
{hu-oconnell.90} The experimental observation of\smallskip $\ \Delta \omega
, $ should be very interesting by evidencing the role of Coulomb effects in
the collective electron motion along the $z$ direction. Note that for the 2D
plasma parameter $\Gamma =\widetilde{e}^{2}/aT\lesssim 1$, where RPA is
formally valid, $\omega _{2D}^{(sh)}\ll \omega _{21}$ and $\Delta \omega
\simeq \omega _{2D}^{(sh)},$ {\it i.e}. the absorption edge is very close to 
$\omega _{21}$ being only slightly shifted to higher frequencies. By
increasing $q$, $\omega _{t}(q)$ decreases according the last term in
brackets in Eq. (\ref{16a}). However our estimates show that, for $T\sim
0.1-1.0$ K and $q\sim 10-10^{2}$ cm$^{-1}$, the coefficient of the quadratic
term is larger than that of linear one. However, in the long wavelength
limit, these coefficients are so small that $\omega _{t}(q)$ $\simeq \omega
_{21}\sqrt{1+2\omega _{2D}^{(sh)}/\omega _{21}}$. As in the longitudinal
mode, $\gamma _{t}(q),$ given by Eq. (\ref{16b}), is exponentially small
such that $\left| \gamma _{t}(q)\right| \ll \omega _{t}(q)$.

The absorption edge of the mode $\omega _{t}(q)$ depends strongly on $\omega
_{21}$. For very small $E_{\perp },$ the experimental values of $\omega
_{21} $ are close to $3\Delta _{0}/4\hbar $ and increase linearly with $%
E_{\perp }. $ \cite{gr-brown.74} For arbitrary $E_{\perp },$ $\omega _{21}$
is obtained from the gap energy, displayed in Fig. 2, and is in the range of
100 GHz to 1 THz. The polarization shift $\Delta \omega \sim \omega
_{sh}^{(2D)}\sim n_{s}$ for $\Gamma <1$, even though $\left| \Delta \omega
\right| \ll \omega _{21}$. For example for $n_{s}=10^{6}$ cm$^{-2}$ and $%
E_{\perp }=0$, we estimate $\omega _{sh}^{(2D)}\sim 100$ MHz $\ll \omega
_{21}$. This makes very difficult the direct experimental observation of
depolarization shift at this electron density. However, the effect should be
observable at higher densities (for instance $n_{s}\sim 10^{8}$ cm$^{-2}$)
even though our results can not be quite reliable in this regime since RPA
should not be valid in such a strongly correlated classical Q2DES. However,
one can hope that the nature of plasma oscillations does not change
drastically, at least qualitatively, even in the high density limit ($\Gamma
>1$) and the depolarization shift should be measured for the Q2DES on the
helium surface.

\subsection{Q1DES}

We now consider plasma oscillations in the Q1DES created along a channel
filled with liquid helium. As in previous work, \cite
{sok-stud.98,kir-mon-kov-grig.93} we consider a parabolic confinement $%
U(y)=m\omega _{0}^{2}y^{2}/2$ with the frequency $\omega _{0}=(eE_{\perp
}mR)^{1/2}$, where $R$ is the curvature radius of the liquid in the channel.
Typical values of $R$ vary from $10^{-4}$ to $10^{-3}$ cm. \cite{kov-mon.86}
The spectrum of electron subbands along the $y$ axis is $E_{n}=\hbar \omega
_{0}(n-1/2),$ $n=1,2,3...$ in addition to the subbands along the $z$
direction. The motion along the $x$ direction (the channel axis) is free.
The frequency $\omega _{0}$ increases with $E_{\perp }$ achieving $100$ GHz
at $E_{\perp }=3$ kV/cm for $R=5\times 10^{-4}$ cm. As $\omega _{0}\ll
\omega _{21}$, the multisubband system in transverse directions can be
decoupled and we ignore electron transitions in $z$ direction which are the
same as discussed above.

The noninteracting density-density response function was calculated in Ref. 
\cite{sok-stud.98}. The result was

\end{mathletters}
\begin{equation}
\Pi _{nn^{\prime }}^{0}(\omega ,q_{x})=-\frac{2N\left[ \exp \left[
-(n-1)\hbar \omega _{0}/T\right] U\left( \zeta _{nn^{\prime }}^{(-)}\right)
-\exp \left[ -(n^{\prime }-1)\hbar \omega _{0}/T\right] U\left( \zeta
_{nn^{\prime }}^{(+)}\right) \right] }{\hbar q_{x}u_{T}\left[ 1+\coth \left(
\hbar \omega _{0}/2T\right) \right] }  \label{16}
\end{equation}
where $\zeta _{nn^{\prime }}^{(\pm )}=(\omega /q_{x}u_{T})\left[ 1+(\omega
_{0}/\omega )(n-n^{\prime })\right] \pm \hbar q_{x}/2mu_{T}$. For $\hbar
\omega _{0}\gg T$, when only the lowest subband ($n=1)$ is occupied, the
expression for the response function is greatly simplified yielding $\Pi
_{nn^{\prime }}^{0}(\omega ,q_{x})=-(N/T)W(\omega /q_{x}u_{T}).$

Using the wave functions of the two lowest subbands $(n=1$ and $n=2)$

\[
\varphi _{1}(y)=\frac{1}{\pi ^{1/4}y_{0}^{1/2}}\exp \left( -\frac{y^{2}}{%
2y_{0}^{2}}\right) ;\qquad \varphi _{2}(y)=\frac{\sqrt{2}}{\pi
^{1/4}y_{0}^{3/2}}y\exp \left( -\frac{y^{2}}{2y_{0}^{2}}\right) , 
\]
where $y_{0}=(\hbar /m\omega _{0})^{1/2}$, we obtain the matrix elements of
Coulomb interaction from Eq. (\ref{1.3}): 
\begin{mathletters}
\begin{equation}
V_{11,11}^{1D}(q_{x})=\frac{\widetilde{e}^{2}}{L_{x}}\exp \left( \frac{%
q_{x}^{2}y_{0}^{2}}{4}\right) K_{0}\left( \frac{q_{x}^{2}y_{0}^{2}}{4}%
\right) \simeq \frac{\widetilde{e}^{2}}{L_{x}}\ln \frac{1}{|q_{x}y_{0}|^{2}}%
\text{ for }|q_{x}y_{0}|\ll 1  \label{19a}
\end{equation}
and 
\begin{equation}
V_{12,12}^{1D}(q_{x})=\frac{\widetilde{e}^{2}}{2L_{x}}\exp \left( \frac{%
q_{x}^{2}y_{0}^{2}}{4}\right) \left[ K_{0}\left( \frac{q_{x}^{2}y_{0}^{2}}{4}%
\right) -\frac{\sqrt{\pi }}{\sqrt{2}|q_{x}y_{0}|}W_{-1,0}\left( \frac{%
q_{x}^{2}y_{0}^{2}}{2}\right) \right]  \label{19b}
\end{equation}
\end{mathletters}
\[
\simeq \frac{\widetilde{e}^{2}}{L_{x}}\left[ 1-\frac{q_{x}^{2}y_{0}^{2}}{2}%
\ln \frac{1}{|q_{x}y_{0}|}\right] \text{ for }|q_{x}y_{0}|\ll 1. 
\]
where $W_{\alpha ,\beta }(x)$ is Whittacker function.

Using Eqs. (\ref{2}), (\ref{3a}), and (\ref{19a}), taking $\omega =\omega
(q_{x})-i\gamma _{q},$ and assuming $\omega (q_{x})/q_{x}u_{T}\gg 1$ and $%
|\omega (q_{x})-\omega _{0}|/q_{x}u_{T}$ $\gg 1$ we obtain the dispersion
relation of the longitudinal intrasubband modes in the long wavelength
limit, $|q_{x}y_{0}|\ll 1$, as 
\begin{mathletters}
\begin{equation}
\omega _{l}(q_{x})=\frac{2\widetilde{e}^{2}q_{x}^{2}}{m\ell }\ln \frac{1}{%
|q_{x}y_{0}|}\exp \left( \frac{q_{x}^{2}y_{0}^{2}}{4}\right) \left[ 1+\frac{%
3T\ell }{2\widetilde{e}^{2}}\ln ^{-1}\frac{1}{|q_{x}y_{0}|}\right] ,
\label{20a}
\end{equation}

\begin{equation}
\gamma _{l}(q_{x})=\sqrt{\pi }\frac{\omega _{l}^{4}(q_{x})}{\left(
q_{x}u_{T}\right) ^{3}}\exp \left[ -\frac{\omega _{l}^{2}(q_{x})}{%
(qu_{T})^{2}}\right] ,  \label{20b}
\end{equation}
where $\ell \simeq n_{l}^{-1}=(N/L_{x})^{-1}$ is the mean interelectron
distance along the channel.

The longitudinal spectrum mode, given by Eq. (\ref{20a}), has the same
structure of the obtained previously in Ref. \cite{sok-stud.98} and in Refs. 
\cite{chaplik,sok-kir.94} where a quasi-crystalline approximation was
employed. However we found an additional second term in brackets, which
should be quite small for reasonable values of $T$ and $\ell $. Note also
that condition $\omega /q_{x}u_{T}\gg 1$ assumed here is equivalent to $T\ll
e^{2}/\ell $ in the quasicrystalline approximation. It worth emphasizes that
the present result was obtained within RPA which is valid in the opposite
limit $T\gg e^{2}/\ell $. Our conclusion is that the plasmon spectrum in the
classical Q1DES has little dependence on the plasma parameter and RPA
results should be probably correct in wide range of electron densities.

The transverse branch of collective excitations is rather interesting.
Following the same steps as before, we arrive to 
\end{mathletters}
\begin{mathletters}
\begin{equation}
\omega _{t}^{2}(q_{x})=\omega _{0}^{2}-\frac{\widetilde{e}^{2}q_{x}^{2}}{%
m\ell }\ln \frac{1}{|q_{x}y_{0}|}  \label{22a}
\end{equation}

\[
+2\omega _{0}\omega _{1D}^{(sh)}\left[ 1+\left( \frac{\left( 1+3\omega
_{1D}^{(sh)}/2\omega _{0}\right) T}{m\left( \omega _{1D}^{(sh)}\right) ^{2}}+%
\frac{\left( 1+\omega _{1D}^{(sh)}/\omega _{0}\right) \hbar }{2m\omega
_{1D}^{(sh)}}\right) q_{x}^{2}\right] , 
\]

\begin{equation}
\gamma _{t}(q_{x})=\sqrt{\pi }\frac{\left[ \omega _{t}(q_{x})-\omega _{21}%
\right] ^{2}}{q_{x}u_{t}}\exp \left[ -\frac{\left[ \omega _{t}(q_{x})-\omega
_{21}\right] ^{2}}{(q_{x}u_{t})^{2}}\right]  \label{22b}
\end{equation}
Here $\omega _{1D}^{(sh)}=\widetilde{e}^{2}/\hbar \ell $. The first two
terms in Eq. (\ref{22a}) correspond to the result obtained in the
quasicrystalline approximation \cite{chaplik,sok-kir.94} if $y_{0}$ is
replaced by $\ell $ in the logarithmic factor. The next term is the
depolarization shift correction increasing the absorption edge frequency by $%
\Delta \omega =\sqrt{\omega _{0}^{2}+2\omega _{0}\omega _{1D}^{(sh)}}-\omega
_{0}\simeq $ $\omega _{1D}^{(sh)}$ when $\omega _{1D}^{(sh)}\ll \omega _{0}$%
. One can see that the instability of the transverse mode ($\omega
_{t}^{2}(q_{x})<0$) in the limit of zero confinement ($\omega _{0}=0$) is
still manifested in our treatment. We call the attention that we found a
quite different result in our previous work \cite{sok-stud.98} because we
used an approximate expression $V_{12,12}^{1D}(q_{x})\simeq \widetilde{e}%
^{2}/L_{x}$ considered in Ref. \cite{hu-oconnell.90}. One estimative is that
the polarization shift correction should be quite small for $n_{l}\sim
10^{2}-10^{3}$ cm$^{-1}$ and $T\simeq 10^{-1}-1$ K such that $e^{2}/\ell <T$%
. For instance, $\omega _{1D}^{(sh)}\simeq 10$ GHz for $E_{\perp }=3$ kV/cm
and $n_{l}=10^{2}$ cm$^{-1}$ whereas $\omega _{0}=100$ GHz at $R=5\times
10^{-4}$ cm. However, this density range can not be achieved in experimental
conditions. For higher densities, $\Delta \omega $ should be of the same
order of $\omega _{0}$ and the polarization shift should be observed even
our results are based on the RPA.

\section{Concluding remarks}

In the present work, we have used the many-body dielectric formalism to
calculate the spectrum of plasma oscillations for the classical Q2DES and
Q1DES formed on the liquid helium surface. We have obtained the general
expression for the density-density Q2D and Q1D response functions for any
frequency and wave number within the RPA. The results are valid at low
temperatures since we have used a two-subband model in which only the lowest
subbands of the motion in the direction normal to the electron layer (Q2D)
and of the motion in direction across the conducting channel (Q1D) are
occupied. The plasma dispersion relations were found from the zeros of the
determinantal equation for the generalized multisubband dielectric
functions. We have obtained corrections to the gapless longitudinal modes,
beyond the q$^{1/2}$-behavior in the Q2DES and sound-like behavior, within
logarithmic accuracy, in the Q1DES. The intersubband transverse collective
frequency is higher than the corresponding single-particle excitation
frequency both in Q2DES and Q1DES. The absorption edge frequencies are
increased by the depolarization shift which can be large at high densities.
In this connection, the experimental study of intersubband transition in
low-dimensional electron systems over liquid helium seems to be attractive,
because of the accessibility of wide range of charge concentrations and low
temperatures, to observe collective effects on spectroscopic transitions. 
\cite{gr-brown.74}

We conclude by pointing out some limitations of our approach. The results
are based on the RPA, which works quite well at small values of the plasma
parameter. We know that RPA results become worse as the dimensionality is
reduced, but we do not know how to go beyond RPA in a controlled way mainly
in the Q1DES. We are, however, encouraged by the good agreement of our RPA
results for the mode spectrum and those obtained in the quasi-crystalline
approximation that is valid in the opposite limit of high plasma parameter.
Other fact is the excellent agreement between the RPA theory and experiment
on collective excitations in semiconductor quantum wells \cite{hansen} and
wires \cite{ulrichs}. Our use of a two-subband model can be and should be
improved in more realistic calculations \cite{sokolov95}. But we do not
expect the correction of including other subbands to be qualitatively
significant though at low temperatures.

\newpage

\begin{center}
{\bf FIGURE CAPTIONS}

\bigskip
\end{center}

\noindent Fig. 1. Variational parameters $\eta _{1}$ (straight line) and $%
\eta _{2}$ (dashed line) as a function of the pressing electric field $%
E_{\bot }$ evaluated numerically from Eqs. (\ref{8a}) and (\ref{8b}).

\bigskip

\noindent Fig. 2. Lowest-subband energy (straight line) and energy gap $%
\Delta _{21}$ (dashed line) of single-electron spectroscopic excitation of
the Q2DES as a function of the pressing electric field $E_{\bot }$.
\end{mathletters}

\end{document}